\documentclass[]{taira}

\usepackage{amssymb}
\usepackage{amsmath}
\usepackage{graphicx}
\usepackage{lipsum}

\definecolor{blue}{rgb}{0, 0.4470, 0.7410}
\definecolor{blue2}{rgb}{0, 0.4470, 0.7410}
\definecolor{blue3}{rgb}{0.1804, 0.0784, 0.5529}
\definecolor{red}{rgb}{0.8500, 0.1250, 0.0480} 
\definecolor{red2}{rgb}{0.8500, 0.1250, 0.0480} 
\definecolor{orange2}{rgb}{0.8500, 0.3250, 0.0980} 
\definecolor{yellow2}{rgb}{0.9290, 0.6940, 0.1250}
\definecolor{purple}{rgb}{0.4940, 0.1840, 0.5560}
\definecolor{purple2}{rgb}{0.4940, 0.1840, 0.5560}
\definecolor{green}{rgb}{0.4660, 0.6740, 0.1880}
\definecolor{green2}{rgb}{0.4660, 0.6740, 0.1880}
\definecolor{ltblue2}{rgb}{0.3010, 0.7450, 0.9330}
\definecolor{dkred2}{rgb}{0.6350, 0.0780, 0.1840}
\definecolor{gray2}{rgb}{0.22, 0.22, 0.3}
\definecolor{gray3}{rgb}{0.5, 0.5, 0.5}

\begin{document}

\shorttitle{Identifying vortical network connectors for turbulent flow modification}
\shortauthor{M. Gopalakrishnan Meena and K. Taira}

\title{Identifying vortical network connectors for turbulent flow modification}

\author{Muralikrishnan Gopalakrishnan Meena\aff{1}\corresp{\email{muraligm@g.ucla.edu}}
 \and Kunihiko Taira\aff{1}
  }

\affiliation{
\aff{1}Department of Mechanical and Aerospace Engineering, University of California, Los Angeles, CA 90095, USA\\
}

\maketitle

\begin{abstract}

We introduce a network (graph) theoretic community-based framework to extract vortical structures that serve the role of connectors for the vortical interactions in two- and three-dimensional isotropic turbulence. The present framework represents the vortical interactions on a network, where the vortical elements are viewed as the nodes and the vortical interactions are regarded as edges weighted by induced velocity from the Biot--Savart law. This formulation enables the use of circulation and spatial arrangement of vortical elements for structure extraction from a flow field. We identify closely interacting vortical elements as vortical network communities through community detection algorithms. We show that the inter- and intra-community interactions can be used to decompose the governing equation for the evolution of network nodes. Furthermore, these community-based interactions are used to identify the communities which have the strongest and weakest interactions amongst them.  These vortical communities are referred to as the connector and peripheral communities, respectively. We demonstrate the influence of the network-based structures to modify the dynamics of a collection of discrete point vortices. Taking advantage of the strong inter-community interactions, connector community can significantly modify the collective dynamics of vortices through the application of multiple impulse perturbations. We then apply the community-based framework to extract influential structures in isotropic turbulence. The connector and peripheral communities extracted from turbulent flows resemble shear-layer and vortex-core like structures, respectively. The influence of the connector structures on the flow field and their neighboring vortical structures is analyzed by adding impulse perturbations to the connectors in direct numerical simulations. The findings are compared with the cases of perturbing the strongest vortex tube and shear-layer regions. We find that perturbing the connector structures enhances local turbulent mixing beyond what are achieved by the other cases.

\end{abstract}

\section{Introduction}

Analysis of turbulence remains as one of the most complex problems in science and engineering due to the strong nonlinear dynamics and multi-scale properties of fluid flows \citep{hussain1986coherent}. As turbulence is ubiquitous in nature and engineering problems, the modification of its dynamics has been an active field of study \citep{brunton2015closed}. For modeling and controlling their dynamics, it is important to understand the interactions amongst the vortical structures.  Insights from such endeavors can support applications including flow separation control \citep{bhattacharjee1986modification} and mixing enhancement \citep{spencer1951mixing}. What makes this control problem challenging is that a large amount of energy is generally required to modify large-scale vortical structures to achieve flow modification. 

To achieve flow modification with low level of energy input, it is critical to identify important vortical structures in the flow. Various techniques have been introduced to extract flow structures. Reduced representation of the flow field using approaches such as the proper orthogonal decomposition (POD; \citet{lumley1967structure}) and dynamic mode decomposition (DMD; \citet{Schmid:JFM10}), are tools which have shown great ability to extract the dominant features of the flow \citep{taira2017modal, taira2020modal}. Measures like \textit{Q}-criterion \citep{Hunt:CTR88}, $\lambda_2$-criterion \citep{Jeong:JFM95}, $\Gamma$-criterion \citep{Graftieaux:MST01}, and finite-time Lyapunov exponent \citep{haller2005objective,haller2015lagrangian}, can be used to identify highly rotational and strained regions of the flow \citep{dubief2000coherent,chakraborty2005relationships}. Recently, machine learning inspired methods have also been used to extract the dominant vortical structures in turbulence \citep{jimenez2018machine}.

Even with the available strategies to identify coherent structures, the quantification and analysis of vortical interactions is a challenge as every element in the flow field interact with others. If the given flow field is spatially discretized into $n$ discrete cells, the number of interactions amongst the cells to be accounted for will be $n(n-1)$. This is particularly crucial in turbulence with high dimensions. Graph theory provides a concrete mathematical framework for representing interactions amongst elements of a system as a network \citep{Bollobas98}. Valuable insights and models for high-dimensional systems, such as the brain networks, have been gained through graph-theoretic formulations \citep{BarabasiNS16}. Moreover, the vast range of tools in network science enables the characterization, modeling, and control of interaction-based dynamics \citep{Newman:10}. 

In recent years, network formulations have been introduced to quantify and capture the interactions in fluid flows. The induced velocity among vortical elements \citep{Nair:JFM15}, Lagrangian motion of fluid elements \citep{ser2015flow,hadjighasem2016spectral}, oscillator-based representation of the energy fluctuations \citep{nair2018networked}, time series of fluid flow properties \citep{scarsoglio2017time}, and triadic interactions in turbulence \citep{gurcan2017nested} have been studied using a network-theoretic framework. The formulations have been extended to characterize various turbulent flows, including two-dimensional isotropic turbulence \citep{Taira:JFM16}, turbulent premixed flames and combustors \citep{singh2017network,godavarthi2017recurrence,krishnan2019emergence}, wall turbulence \citep{iacobello2018visibility}, mixing in turbulent channel flow \citep{iacobello2018spatial,iacobello2019lagrangian}, and isotropic magnetohydrodynamic turbulence \citep{gurcan2018nested,gurcan2020turbulence}. 

Network-based clustering techniques have also been utilized to extract closely connected nodes and dominant features in fluid flows. Image sequences have been used to reconstruct the flow field using the Frobenius--Perron operator and community detection is implemented to identify key structures in the phase space \citep{bollt2001combinatorial}. Spectral clustering has been considered for vortex detection in a Lagrangian-based framework of fluid flow networks \citep{hadjighasem2016spectral}. A coherent structure coloring technique also builds on a Lagrangian framework to identify coherent structures in complex flows \citep{schlueter2017coherent,husic2019simultaneous}. More recently, community detection has been used to extract vortical structures to form reduced-order models for laminar wake flows \citep{Meena:PRE18,Meena2020PhD}.

A key attribute missing in clustering approaches is to take advantage of the inter- and intra-cluster interactions to identify the important interactions and clusters in the flow. Moreover, modifying the system dynamics by taking advantage of the interactions amongst the clusters have not yet been explored. In the present study, we use the intra- and inter-cluster interactions extracted from a network-based framework for identifying important flow-modifying vortical structures. We use the community detection algorithm \citep{Meena:PRE18,Meena2020PhD} to extract closely connected vortical elements in two- and three-dimensional isotropic turbulence. The interactions amongst the communities are used to identify key turbulent flow-modifying structures. The goal of this network-based framework is not to alter the global turbulent flow, but to influence certain key vortical structures in the complex background of isotropic turbulence.

\begin{figure}
  \centerline{\includegraphics[width=0.97\textwidth]{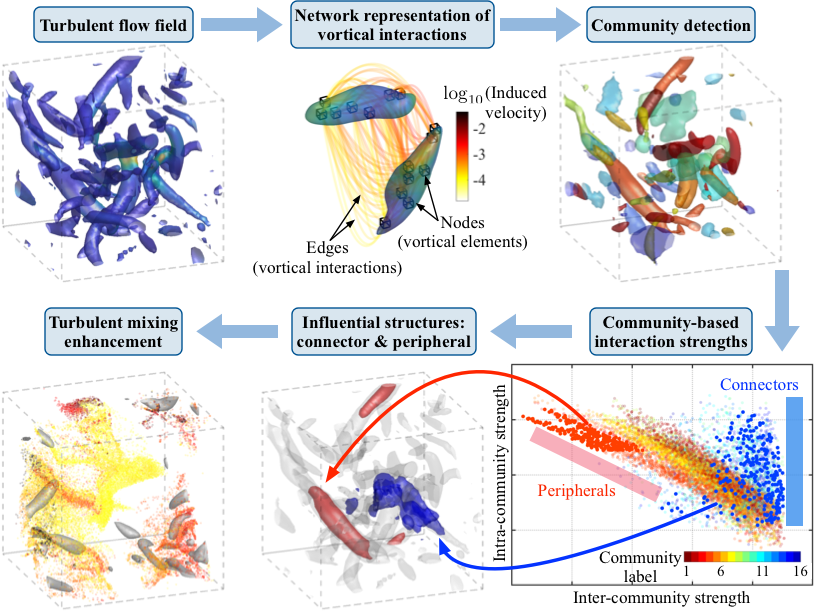}}
  \caption{An overview of the community-based procedure for extracting turbulent flow-modifying structures.}
\label{f:overview}
\end{figure}

A procedure for extracting the community-based structure for turbulent flows is illustrated in figure~\ref{f:overview}. In what follows, we first introduce the network representation of vortical interactions in \S\ref{s:theory_net}. We introduce the network-based measure of node strength and community detection to identify influential nodes in \S\ref{ss:theory_net_extract_str} and \S\ref{ss:theory_net_extract_comm}, respectively. The importance of the network-based measures is discussed within the context of a model problem of ideal point-vortex dynamics. We assess the influence of the identified nodes to modify the dynamics of a collection of discrete point vortices in \S\ref{s:point_vortex}. We then employ the network community-based formulation to extract influential structures in two- and three-dimensional isotropic turbulence, described in \S\ref{s:iso_turb}. The numerical setups are discussed in \S\ref{ss:iso_turb_setup}. We characterize the vortical network of two- and three-dimensional isotropic turbulence in \S\ref{ss:iso_turb_charac}. We then demonstrate the use of community-based structures to modify the turbulent flows in \S\ref{ss:iso_turb_perturbation}. Finally, concluding remarks are provided in \S\ref{s:conclusion}. 

\section{Network-theoretic description of vortical interactions}\label{s:theory_net}

\begin{figure}
  \centerline{\includegraphics[width=0.4\textwidth]{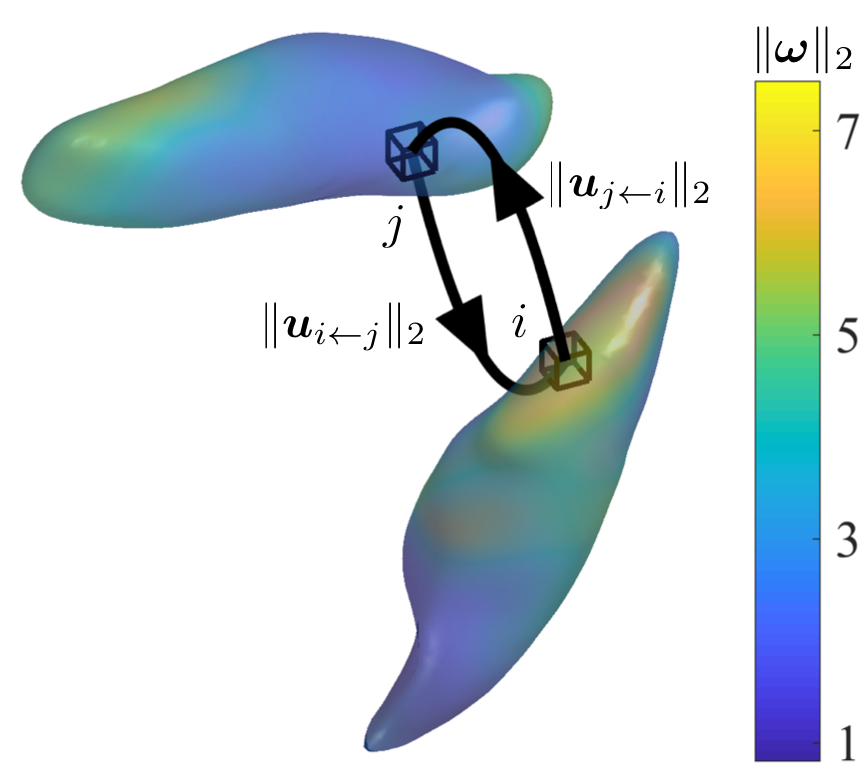}}
  \caption{Interactions between two vortical elements in vortical structures extracted from three-dimensional isotropic turbulent flow. The vortical structures are visualized by isosurface of \textit{Q}-criterion \citep{Hunt:CTR88} coloured with $\|\boldsymbol{\omega}\|_2$. The vortical elements are shown for the spatial grid cells.}
\label{f:initial_net}
\end{figure}

To identify the influential regions to perturb for flow modification, we examine the interactions amongst the vortical elements. We discretize the vorticity field in a Lagrangian and Eulerian perspective. The discrete vortical elements are referred to as nodes in the present work. To quantify the interactions amongst the vortical nodes, we consider the induced velocity imposed upon each other. The Biot--Savart law provides the induced velocity from a vortical element as a function of circulation and relative position of the vortical elements, expressed as
\begin{equation}
    \boldsymbol{u}(\boldsymbol{r},t) 
    = \frac{1}{2(n_\text{d}-1)\pi}
    \int_{V} \frac{\boldsymbol{\omega}(\boldsymbol{r}',t) \times (\boldsymbol{r}-\boldsymbol{r}')}
    {\|\boldsymbol{r}-\boldsymbol{r}'\|_2^{n_\text{d}}} 
    \text{d}V',
\label{e:BiotSavart}
\end{equation}
where $\boldsymbol{u}(\boldsymbol{r},t)$ is the induced velocity at location $\boldsymbol{r}$ in the domain from a collection of vortical elements enclosed in volume $V$ with a vorticity distribution of $\boldsymbol{\omega}(\boldsymbol{r}',t)$ at positions $\boldsymbol{r}'$. Here, $n_\text{d}$ is the spatial dimension of the flow field. Furthermore, the influence of a vortical node (element) $i$ on node $j$ can be written as
\begin{equation}
    \boldsymbol{u}_{j\leftarrow i} = \frac{\Gamma_i \text{d}l}{2(n_\text{d}-1)\pi} \frac{\sin\theta}{\|\boldsymbol{r}_j - \boldsymbol{r}_i\|_2^{n_\text{d}-1}},
\label{e:induced_decouple}
\end{equation}
where $\Gamma = \|\boldsymbol{\omega}(\boldsymbol{r},t)\|_2 \text{d}S$ is the circulation of a vortical element of area $\text{d}S$ and length $\text{d}l$, and $\theta$ is the angle between the vorticity and distance vectors. As an example, we illustrate the interactions between two vortical nodes in figure~\ref{f:initial_net}. Here, element $i$ has higher vorticity magnitude compared to element $j$. Thus, the velocity induced by $i$ onto $j$ is higher than that imposed by $j$ onto $i$, yielding an asymmetric interaction.

Characterizing the interaction-based behavior of vortical elements can be a challenge in high-Reynolds number turbulence, particularly with high degrees of freedom needed to discretize the flows. To facilitate the analysis of high-dimensional dynamics, we leverage the analytical approaches in graph theory \citep{Bollobas98} and network science \citep{Newman:10}. Here, we establish a network-theoretic representation of the vortical interactions in a flow field. We construct a network (graph) $\mathcal{G}$ comprised of vortical nodes $\mathcal{V}$ connected by edges $\mathcal{E}$ holding edge weights $\mathcal{W}$ based on induced velocity. Given this definition $\mathcal{G} = \mathcal{G}(\mathcal{V},\mathcal{E},\mathcal{W})$ for the network, we can quantify the important nodes in vortical flows.

The collection of connectivity amongst the nodes can be represented by the adjacency matrix $\boldsymbol{A}$, which holds the edge weights as its elements. For the vortical interaction network, $\boldsymbol{A}$ can be defined using the normalized induced velocity \citep{Nair:JFM15,Taira:JFM16} as
\begin{equation}
    A_{ij} = \frac{\|\boldsymbol{u}_{i\leftarrow j}\|_2}{u^*},
\label{e:adjacency}
\end{equation}
where $u^*$ is a characteristic velocity of the flow. For a flow field with $n$ vortical nodes, the adjacency matrix $\boldsymbol{A} \in \mathbb{R}^{n\times n}$. The above formulation gives an asymmetric adjacency matrix, representing a directed network. Adding directions to the links helps differentiate between the influential and influenced nodes. Non-dimensionalization is important for the analysis of turbulent flows over a range of Reynolds number. The details of the non-dimensionalization will be discussed in \S\ref{s:iso_turb}.

Let us discuss the role of $\boldsymbol{A}$ on network dynamics and appropriate measures to identify influential nodes. Consider a general dynamical system for $n$ state vectors $\boldsymbol{x}_i \in \mathbb{R}^{p_\text{v}}$ holding $p_\text{v}$ variables over a network. The general interaction-based dynamics of the elements can be expressed as
\begin{equation}
    \dot{\boldsymbol{x}}_i = \boldsymbol{f}(\boldsymbol{x}_i) + \sum_{j=1}^n A_{ij} \boldsymbol{g}(\boldsymbol{x}_i, \boldsymbol{x}_j),
   \quad i = 1, 2, \dots, n,
   \label{e:net_dyn}
\end{equation}
where function $\boldsymbol{f}(\boldsymbol{x}_i)$ represents the intrinsic dynamics of node $i$ and function $\boldsymbol{g}(\boldsymbol{x}_i,\boldsymbol{x}_j)$ describes the interactive dynamics between nodes $i$ and $j$. We consider the model fluid flow problem of ideal point-vortex dynamics \citep{aref2007point,newton2013n} to demonstrate the interaction-based dynamics of vortical elements and identify important vortical nodes using the network-based approach.

\begin{figure}
  \centerline{\includegraphics[width=0.97\textwidth]{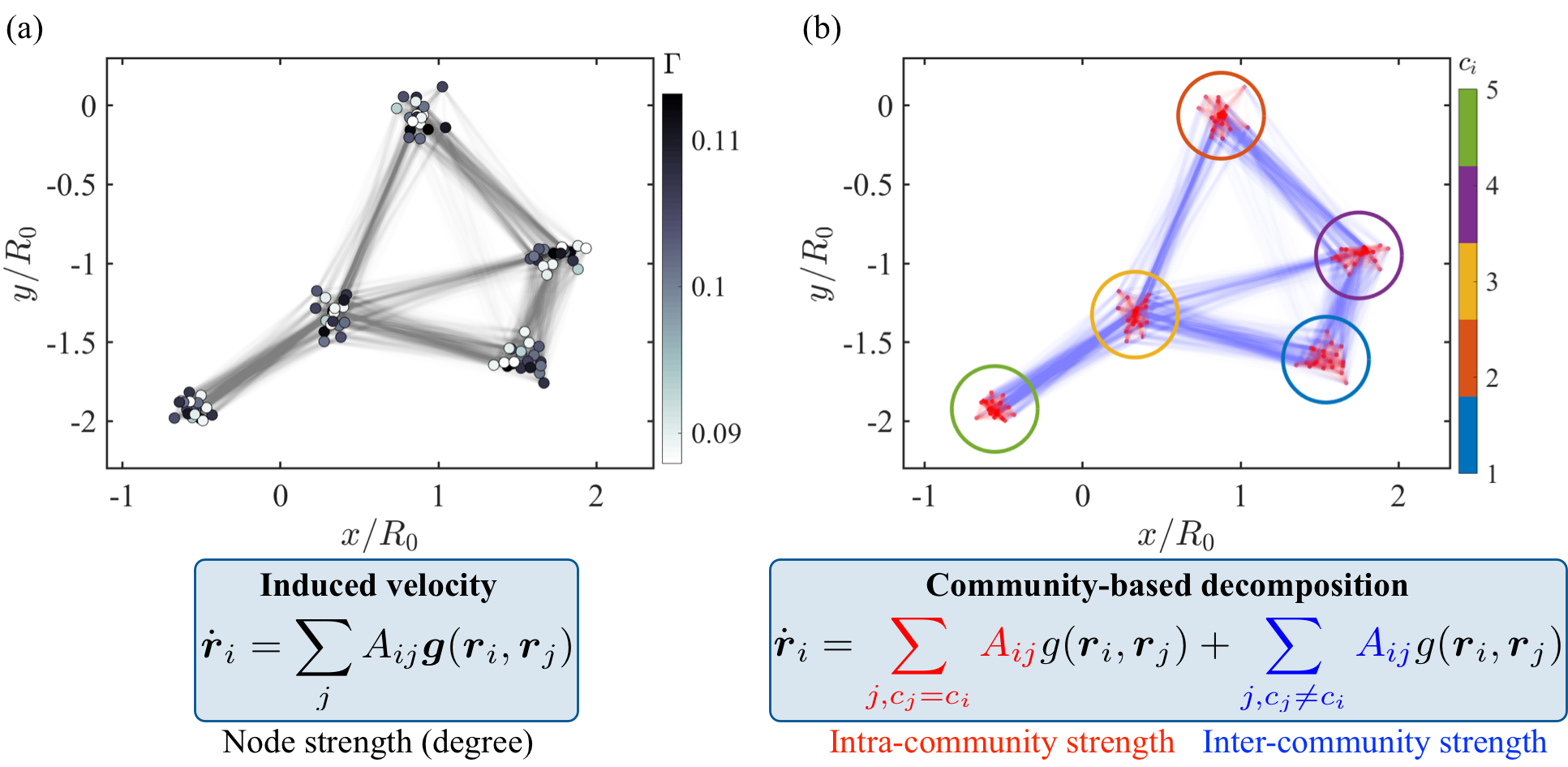}}
  \caption{Interactions amongst a collection of discrete point vortices are used to illustrate the decomposition of networked dynamics through intra- and inter-community interactions.}
\label{f:comm_dynamics}
\end{figure}

Let us take a collection of $n = 100$ discrete point vortices, initialized on an infinite two-dimensional domain as shown in figure~\ref{f:comm_dynamics}~(a). These vortices are arranged into five groups at the initial time. The vortices are coloured by their circulations $\Gamma_i$, which is kept constant over time in the inviscid flow. The circulations have a normal distribution about a mean of $\overline{\Gamma} = 0.1$ and a standard deviation of $\sigma_{\Gamma} = 0.008$. This canonical model problem portrays the nonlinear dynamics of vortical structures found in various flows \citep{Nair:JFM15}. The transparent gray edges visualize all interactions amongst the nodes, based on equation~\ref{e:adjacency}. Here, we use $u^* = \Gamma_{\text{tot}}/(2\pi R_0)$ where $\Gamma_{\text{tot}} = \sum_i^n \Gamma_i$ is the total circulation of the system and $R_0$ is the average radial distance of the centroid of the clusters from the geometric center of the overall system at initial condition. The spatial variables are non-dimensionalized by $R_0$. The Biot--Savart law governs the dynamics of the point vortices, which can be expressed in terms of equation \ref{e:net_dyn} for which $\boldsymbol{f}(\boldsymbol{r}_i) = 0$ and $\boldsymbol{g}(\boldsymbol{r}_i, \boldsymbol{r}_j) = u^* \hat{\boldsymbol{k}}\times(\boldsymbol{r}_i - \boldsymbol{r}_j)/\|\boldsymbol{r}_i - \boldsymbol{r}_j\|_2$. Here, $\hat{\boldsymbol{k}}$ is the out-of-plane unit normal vector. We use this dynamical systems example to demonstrate the evaluation of important network measures pertinent to the interactive dynamics of the vortical elements.

\subsection{Node strength}\label{ss:theory_net_extract_str}

The node strength measures the ability of a node to be influential (out-strength) or be influenced (in-strength) in the network. The out- and in-strengths of a node are defined by
\begin{equation}
    s_i^\text{out} = \sum\limits_{j}A_{ji}
    \quad 
    \text{and}
    \quad 
    s_i^\text{in} = \sum\limits_{j}A_{ij},
\label{e:str_out_in}
\end{equation}
respectively. These measures are useful to identify the node with collection of significant connections in a network.

For the vortical networks, the present definition of edge weight from equation \ref{e:adjacency} is based on the out-strength. Herein, the node strength, $s_i$, is taken to be the out-strength, unless specified. A relation between the node strength and enstrophy $\Omega(\boldsymbol{r},t)$ of a vortical element can be obtained using equation~\ref{e:induced_decouple} as
\begin{equation}
s_i = \frac{\Gamma_i \text{d}l}{2(n_\text{d}-1)\pi} \sum\limits_j \frac{\sin\theta}{\|\boldsymbol{r}_{j}-\boldsymbol{r}_{i}\|_2^{n_\text{d} - 1}} = \frac{C}{2(n_\text{d}-1)\pi} \sqrt{\Omega_i} \text{d}V,
\label{e:str_enstrophy}
\end{equation}
where the sum of distance components $C$ is constant in a fully periodic domain or can be inferred given the location of the nodes. This relationship reveals that $s_i \propto \sqrt{\Omega_i}$. This is particularly useful to determine the node strength distribution $p(s)$ of a vortical network. The distribution is dependent on the enstrophy distribution $p(\Omega)$. The latter is usually a known or measurable flow statistics. Distribution $p(s)$ gives a global picture of the nature of connectivity in the network and is used to identify the type of the network \citep{BarabasiNS16}. The distribution is also useful to identify nodes with high $s$, which is referred to as the hub nodes. These nodes have been found to be important to assess the robustness of the network dynamics against random and targeted perturbations \citep{Albert:Nature00,Taira:JFM16}.

\subsection{Community detection}\label{ss:theory_net_extract_comm}

Identifying closely connected vortical nodes is important towards revealing key local groups on the network. Such modular groups of nodes with high connectivity amongst each other are referred to as communities \citep{newman2004finding}. One approach to find the communities is to measure the overall modular nature of a network using modularity $M$ given by \citep{leicht2008community}
\begin{equation}
   M = \frac{1}{2n_\text{e}}\sum\limits_{ij}\left[ A_{ij} - \gamma_\text{M}\frac{s_i^\text{in}s_j^\text{out}}{2 n_\text{e}} \right]\delta(c_i,c_j),
   \label{e:modularity}
\end{equation}
where $n_\text{e}$ is the total number of edges in the network, $\gamma_\text{M}$  is the modularity resolution parameter to weigh the presence of small or large communities in the network \citep{reichardt2006statistical,fortunato2007resolution}, $\delta(c_i,c_j)$ is the Kronecker delta, $c_i \in \hat{C}_k$ is the label of the community to which element $i$ is assigned and $\hat{C}_k$ is the set of $k$-th network community. Here, $k = 1,2, \dots, m$, with $m$ being the total number of communities. The communities can be identified by maximizing $M$ by regrouping the nodes. Here, the number of communities $m$ is unspecified and determined by the algorithm. Various algorithms are available to identify the communities in a network \citep{Fortunato:PR10}. In the present study, we adopt the method by \cite{blondel2008fast} to identify the communities in large vortical networks with accuracy and low computational cost \citep{Fortunato:PR10}. We herein refer to these network communities on vortical networks as the vortical communities \citep{Meena:PRE18}. Also, for vortical flows, $\gamma_\text{M}$ can be set based on the plateau effect on the number of vortical communities identified with change in $\gamma_\text{M}$ for a given Reynolds number.

The community information can then be used to decompose the term with the adjacency matrix in equation~\ref{e:net_dyn} as
\begin{equation}
    \dot{\boldsymbol{x}}_i = \boldsymbol{f}(\boldsymbol{x}_i) + \sum\limits_{j, c_j = c_i} A_{ij} \boldsymbol{g}(\boldsymbol{x}_i,\boldsymbol{x}_j) + \sum\limits_{j, c_j \ne c_i} A_{ij} \boldsymbol{g}(\boldsymbol{x}_i,\boldsymbol{x}_j).
\label{e:net_dyn_comm}
\end{equation}
The second term on the right-hand side represents the interaction of node $i$ with the nodes in its own community and the third term denotes the interaction of node $i$ with the nodes in the other communities. The former represents the intra-community interactions and the latter term captures the inter-community interactions of node $i$. This gives a community-based dynamical systems equation for the elements of a network, emphasizing local influences of the communities. An illustration of the above procedure applied to the system of point vortices is shown in figure~\ref{f:comm_dynamics}. The network with no distinction of the weighted edges is shown in figure~\ref{f:comm_dynamics}~(a). The community detection algorithm classifies the nodes into several communities, highlighted by the coloured circles in figure~\ref{f:comm_dynamics}~(b). The intra- and inter-community edges are shown in red and blue, respectively.

Let us quantify the local influence of a node using the above formulation. Similar to how the node strength is defined in equation \ref{e:str_out_in}, the strength of node $i$ to influence all nodes in community $k$ can be defined as
\begin{equation}
    s_{i,k} = \sum\limits_{j,c_j \in \hat{C}_k} A_{ji}.
    \label{e:str_comm}
\end{equation}
Moreover, the strength of a node on the network can be separated into intra- and inter-community strengths as
\begin{equation}
    s_{i}^{\text{intra}} = \sum\limits_{j, c_j=c_i} A_{ji} = s_{i,c_i}
    \quad 
    \text{and}
    \quad 
    s_{i}^{\text{inter}} = \sum\limits_{j,c_j\ne c_i} A_{ji} = \sum\limits_{k,k \ne c_i}s_{i,k},
\label{e:str_comm_inter_intra}
\end{equation}
respectively. These community-based strengths can be used to quantify the interactions with respect to communities.

The intra-community strength can be normalized as the within-module \textit{z}-score \citep{guimera2005functional} given by
\begin{equation}
    Z_i = \frac{s_{i}^{\text{intra}} - \overline{s_{i}^{\text{intra}}}}{\sigma_{s_{i}^{\text{intra}}}},
\label{e:zscore}
\end{equation}
where $\overline{s_{i}^{\text{intra}}}$ and $\sigma_{s_{i}^{\text{intra}}}$ are the mean and standard deviation of $s_{i}^{\text{intra}}$ over all nodes in the community of $i$. The within-module \textit{z}-score identifies the most well-connected node inside a community or the hub node of a community. Note that the hub node of a community need not be well-connected with the other communities in the network. 

A relative measure of inter-community strength of a node, quantified by how well-distributed its edges are amongst communities, can be given by the participation coefficient \citep{guimera2005functional}
\begin{equation}
    P_i = 1 - \left[ \left( \frac{s_{i}^{\text{intra}}}{s_i} \right) ^2 + \sum\limits_{k,k \ne c_i}\left( \frac{s_{i,k}}{s_i} \right) ^2 \right].
\label{e:participation}
\end{equation}
When $P_i\approx 1$, edge weights of node $i$ are equally connected among all communities and $P_i=0$ when the node is only connected to its own community. Note that $s_{i,k}$, $Z_i$, and $P_i$ can be evaluated for both in- and out-edges. In the present study, we evaluate the out-edge based measures following the definition of edge weight from equation~\ref{e:adjacency}.

\begin{figure}
  \centerline{\includegraphics[width=0.97\textwidth]{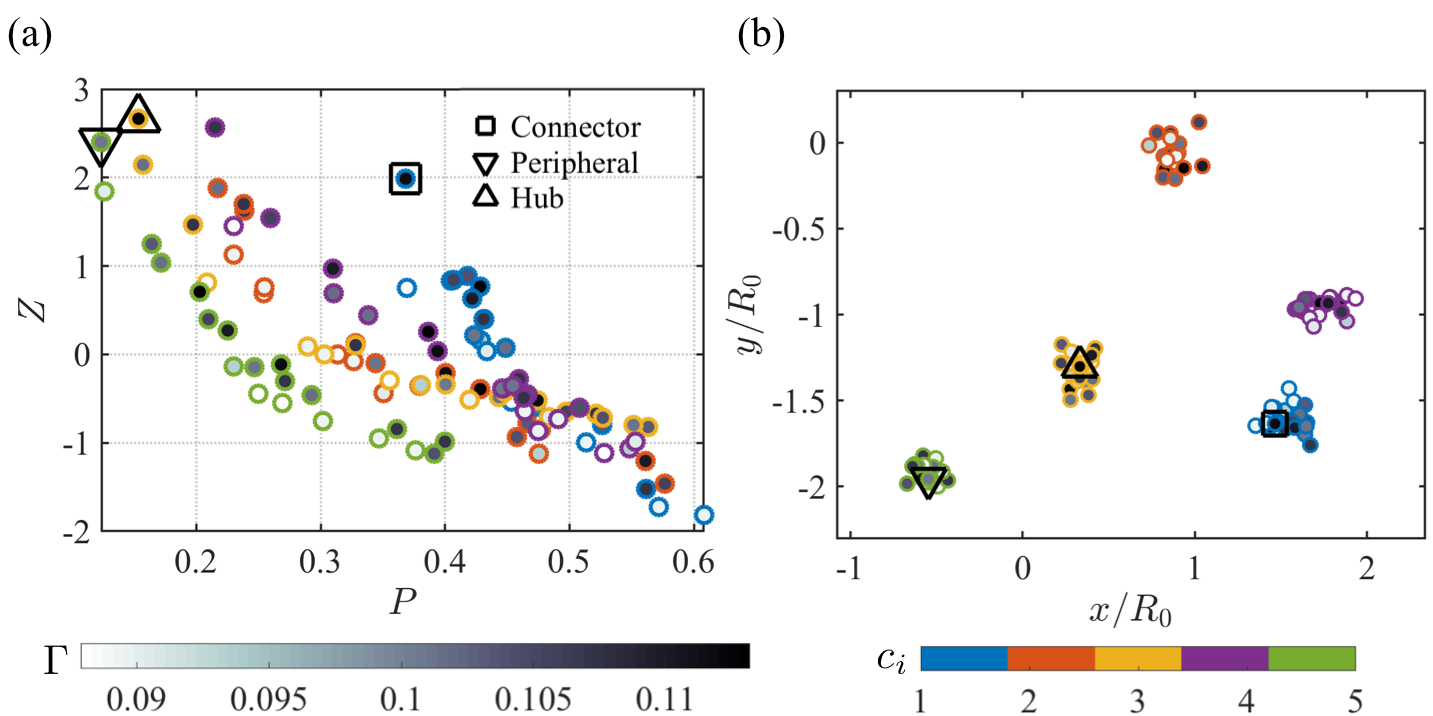}}
  \caption{(a) Distribution of the discrete point vortices in the $P-Z$ map. (b) Position of the important nodes in physical space.
  }
\label{f:Pz_opt}
\end{figure}

The $P-Z$ map of the nodes is an important feature space for revealing key elements within and amongst the communities \citep{guimera2005worldwide,rubinov2010complex,Fortunato:PR10}. The $P-Z$ map for the system of point vortices is shown in figure~\ref{f:Pz_opt}~(a). The nodes on the left side of the $P-Z$ map have very low inter-community interactions and are isolated groups, called \textit{peripherals}. On the other hand, the nodes on the right side have high inter-community interactions making them well-connected to most communities and are called \textit{connectors}. The nodes on the top region with high $Z$ value have the strongest interactions within the respective communities. Note that the name peripheral need not refer to nodes at the physical perimeter of the domain. The strongest peripheral nodes will have high influence within their communities but the least influence on other communities. In contrast, the strongest connector nodes will have the highest influence amongst communities, particularly on the neighbouring communities, and high influence within their own community.

We use the $P-Z$ feature space to search for the strongest peripheral and connector nodes in a network, herein referred to as simply peripheral and connector nodes. We find the average $P_i$ of each community $k$, denoted as $\overline{P_k}$. The peripheral node of a network is the node $i$ given by $\max_{i,c_i=k}Z_i$, belonging to the community $k$ with $\min_{k}\overline{P_k}$. The connector node $i$ is given by $\max_{i,c_i=k}Z_i$, belonging to community $k$ with $\max_{k}\overline{P_k}$. The connector, peripheral, and hub nodes of the discrete point vortex system are indicated in figure~\ref{f:Pz_opt}~(a). The $P-Z$ values for the nodes suggest that the hub node would have similar characteristics as the peripheral node. The three important nodes are also highlighted in the physical space as shown in figure~\ref{f:Pz_opt}~(b). The positions of the peripheral and connector nodes suggest that an influential node need not be located at the geometric center of the community. The observations signify the need to consider inter- and intra-community interactions for identifying important nodes. Let us now demonstrate how the behavior of the networked system can be modified using these measures.
\section{Community-based modification of discrete point-vortex dynamics}\label{s:point_vortex}

We analyze the influence of the connector, peripheral, and hub nodes identified using the network-based measures on the dynamics of a collection of discrete point vortices. We consider the same model problem setup with $n=100$ discrete point vortices used in \S\ref{s:theory_net}. Velocity based impulse perturbations are added to the communities corresponding to the influential nodes. Based on the nodes, we refer to the perturbations as connector, peripheral, and hub-based perturbations. Positions of the perturbed communities at initial time are shown in figure~\ref{f:Pz_opt}~(b). We identify the influential nodes only at initial time as we are interested to explore the influence of the nodes on the system dynamics. 

\begin{figure}
  \centerline{\includegraphics[width=0.55\textwidth]{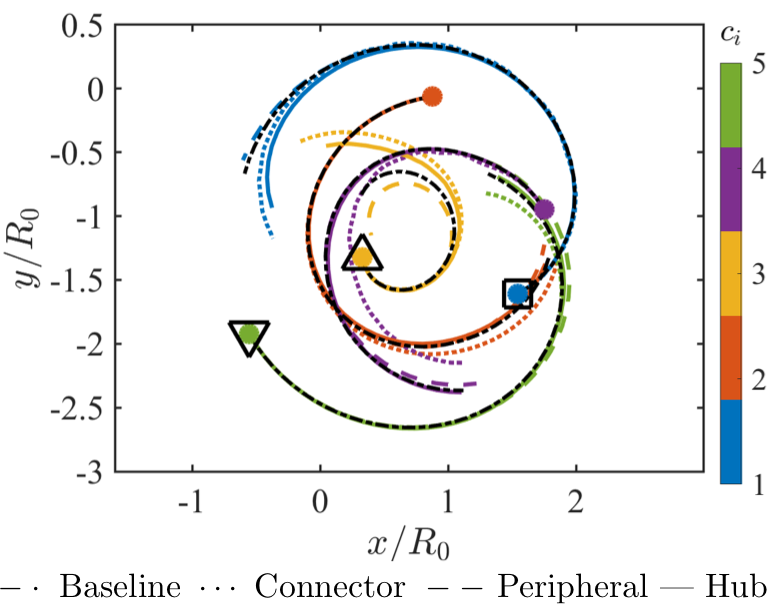}}
  \caption{Trajectories of community centroids without and with connector ($\square$), peripheral ($\bigtriangledown$), and hub-based ($\bigtriangleup$) perturbations. Filled circles show initial position of the community centroids.}
\label{f:2Dpoint_results1}
\end{figure}

Impulse perturbations at discrete time $n_\text{t}\Delta t$ are added to the velocity field with time step $\Delta t$ and $n_\text{t} = 0,1,2,\dots$. The velocity of a perturbed node $i$ at time $t$ is given by $\boldsymbol{u}(\boldsymbol{r}_i,t) + \tilde{\boldsymbol{u}}(\boldsymbol{r}_i,t)$, where $\tilde{\boldsymbol{u}}(\boldsymbol{r}_i,t) = \alpha\hat{e}_{\boldsymbol{u}(\boldsymbol{r}_i,t)} \delta(t-n_\text{t}\Delta t)$, $\alpha$ is the amplitude of perturbation, and $\hat{e}_{\boldsymbol{u}(\boldsymbol{r}_i,t)}$ is the unit vector in the direction of $\boldsymbol{u}(\boldsymbol{r}_i,t)$. Time is non-dimensionalized as $t\Gamma_{\text{tot}}/(2\pi R_0^2)$. Amplitude of perturbation $\alpha$ is computed for a given energy ratio $E$ of
\begin{equation}
    E = \frac{\sum\limits_{i,c_i\in \hat{C}_p} \| \tilde{\boldsymbol{u}}(\boldsymbol{r}_i,t) \|_2^2}{\sum\limits_i^n\| \boldsymbol{u}(\boldsymbol{r}_i,t)\|_2^2},
\end{equation}
where $\hat{C}_p$ is the set of the perturbed community. We have analyzed the system dynamics with $E$ varied between $0.01-0.1$ and have found qualitative similarity in the results. Here, we show the results of perturbations with $E=0.1$ to portray significant changes in the vortex trajectories. The time step between each perturbation is $\Delta t\Gamma_{\text{tot}}/(2\pi R_0^2) = 0.024$.

The effect of the perturbations on system dynamics is assessed by observing the change in trajectories of the community centroids, as shown in figure~\ref{f:2Dpoint_results1}. The centroid location $\boldsymbol{\xi}_k(t)$ of each community $k$ is computed as 
\begin{equation}
    \boldsymbol{\xi}_k(t) = \frac{\sum_{i,c_i\in \hat{C}_k} \Gamma_i \boldsymbol{r}_i(t)}{\sum_{i,c_i\in \hat{C}_k}\Gamma_i}.
\end{equation}
The connector-based perturbations achieve the largest deviations on the trajectories. The peripheral and hub-based perturbations have significant influence only on their neighbouring communities, communities $3$ and $1$, respectively. Regardless of the central location of the hub community at initial time, the system dynamics is not changed compared to the extent achieved by connector-based perturbations. This demonstration highlights the need to consider the strengths and relative positions in a systematic manner to identify the influential nodes.

\begin{figure}
  \centerline{\includegraphics[width=0.97\textwidth]{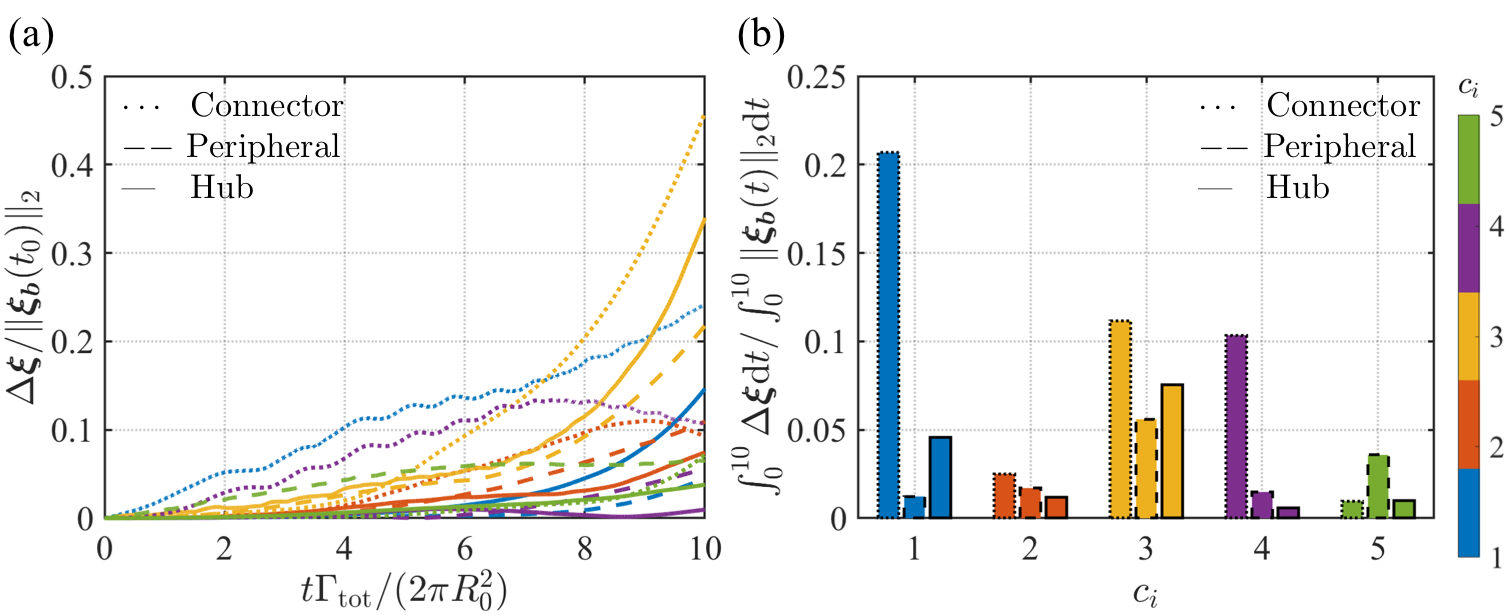}}
  \caption{(a)~Trajectories of community centroids subjected to perturbations compared with baseline, $\boldsymbol{\xi}_b$. Here, $\Delta \boldsymbol{\xi} = \| \boldsymbol{\xi}(t) - \boldsymbol{\xi}_b(t) \|_2$. (b)~Total change in trajectory of community centroids with respect to the baseline.}
\label{f:2Dpoint_results2}
\end{figure}

Let us compare the community centroid trajectories in time, as shown in figure~\ref{f:2Dpoint_results2}~(a). The trajectories of all other communities show the most deviation from baseline with connector-based perturbation, except for community~5. The trajectory of community~5, the peripheral community, is modified significantly when it is perturbed. We have evaluated the $P-Z$ map using the in-edges (not shown), which also gives community~5 to be the peripheral community. Thus, other communities have less influence on community~5. We concentrate on the results at early times to assess the characteristics of the connector to influence its neighbour and connect with other communities. The trajectory of community $4$ is changed significantly earlier in time with connector-based perturbations. This change in the trajectory of communities~4, the spatially closest community to the connector, at early times demonstrates the ability of a connector community to significantly influence its neighbour. Neither of the peripheral and hub-based perturbations influence other communities, particularly their respective neighbours, at early times. Later, the connector-based perturbations also significantly change the dynamics of community $3$ by connecting through community $4$, even though communities $1$ and $3$ are spatially far apart. The inter-community influence demonstrates the connecting characteristics of a connector community.

Deviation of the trajectories from baseline is quantified in figure~\ref{f:2Dpoint_results2}~(b). The observations quantitatively show the larger deviations in trajectories resulting from connector-based perturbations. Considering the neighbours of the perturbed communities, a total change between $10-12\%$ of the baseline is achieved for communities $3$ and $4$ using connector-based perturbations. The corresponding changes using peripheral and hub-based perturbations are around $5\%$ for communities $3$ and $1$, respectively. The largest effect of the hub community is on its own trajectory. These observations demonstrate the inference from the $P-Z$ map that the hub community can portray characteristics of a peripheral. The magnitude of change achieved by the connector-based perturbation on its own trajectory is more than thrice that produced by peripheral and hub communities. Furthermore, the total change in trajectory of community $3$ is the highest using connector-based perturbation. For this model problem, we have demonstrated that the connector node effectively modifies the global vortex dynamics. The present finding motivates the use of inter-community interactions to instigate changes in the global system dynamics, based on the community-based decomposition of system dynamics to identify key structures. 
\section{Community-based flow modification of isotropic turbulence}\label{s:iso_turb}

Let us now consider the application of community-based flow modification to two- and three-dimensional decaying homogeneous isotropic turbulence. The highly complex and multi-scale properties of isotropic turbulence make it an apt choice to demonstrate the capability of the present community-based framework. Isotropic turbulence is a canonical model problem for a range of turbulent flows encountered in nature and engineering applications. 

\subsection{Numerical setup}\label{ss:iso_turb_setup}

For the two- and three-dimensional isotropic turbulent flows, we use the Fourier spectral and pseudo-spectral algorithms, respectively, to numerically solve the Navier-Stokes equations \citep{Taira:JFM16,chumakov2008priori}. Direct numerical simulations (DNS) of the flows are performed in bi-periodic and tri-periodic square and cubic domains of length $L$. For the simulations, the flow fields are resolved such that $k_{\text{max}}\eta \ge 1$, where $k_{\text{max}}$ is the maximum resolvable wavenumber of the grid and $\eta$ is the Kolmogorov length scale. We non-dimensionalize the spatial variables by $L$, time by the large eddy turn-over time at initial time $\tau_e(t_0)$.

The two-dimensional turbulent flows with an initial Taylor microscale based Reynolds number of $Re_\lambda(t_0) \approx 4000$ are obtained from DNS performed at a grid resolution of $1024\times 1024$. We use snapshots of the vorticity field, uniformly sub-sampled to a resolution of $128\times 128$, to construct the vortical network. For three-dimensional isotropic turbulence, flow fields with $Re_\lambda(t_0) \approx 40$ are obtained from DNS performed with a grid resolution of $64 \times 64 \times 64$. The three-dimensional flow fields are uniformly sub-sampled to a resolution of $32\times32\times32$ for constructing the vortical network. Sub-sampling is performed in a manner such that the network representation is not influenced.

The network representation can be made independent of the Reynolds number following the non-dimensionalization of the edge weights using equation~\ref{e:adjacency}. We choose the characteristic velocity 
\begin{equation}
    u^* = V^{1/n_\text{d}}\Omega_{\text{tot}}^{1/2} = V^{1/n_\text{d}} \left(\frac{\int_{\|\boldsymbol{\omega}\|_2\ge\omega_{\text{th}}} \|\boldsymbol{\omega}(\boldsymbol{r},t)\|_2^2 \text{ d}V}{V}\right)^{1/2},
\end{equation}
where $\Omega_{\text{tot}}$ is the total enstrophy per unit area or volume of all the vortical elements enclosed in a region of vorticity threshold $\omega_\text{th}$. We concentrate on vortical elements with high vorticity \citep{mcwilliams1984emergence}, extracted through vorticity thresholding \citep{jimenez1993structure}. We can capture the overall interaction behavior of the flow field even with the threshold. For both two- and three-dimensional flow fields, we use a threshold of $\omega_\text{th}=0.05||\boldsymbol{\omega}(\boldsymbol{r})||_\infty$. Detailed assessment on the influence of the Reynolds number, grid, and $\omega_\text{th}$ is provided in Appendix \ref{app:str_grid_Re}.

\subsection{Network characterization of isotropic turbulence}\label{ss:iso_turb_charac}

\begin{figure}
  \centerline{\includegraphics[width=0.97\textwidth]{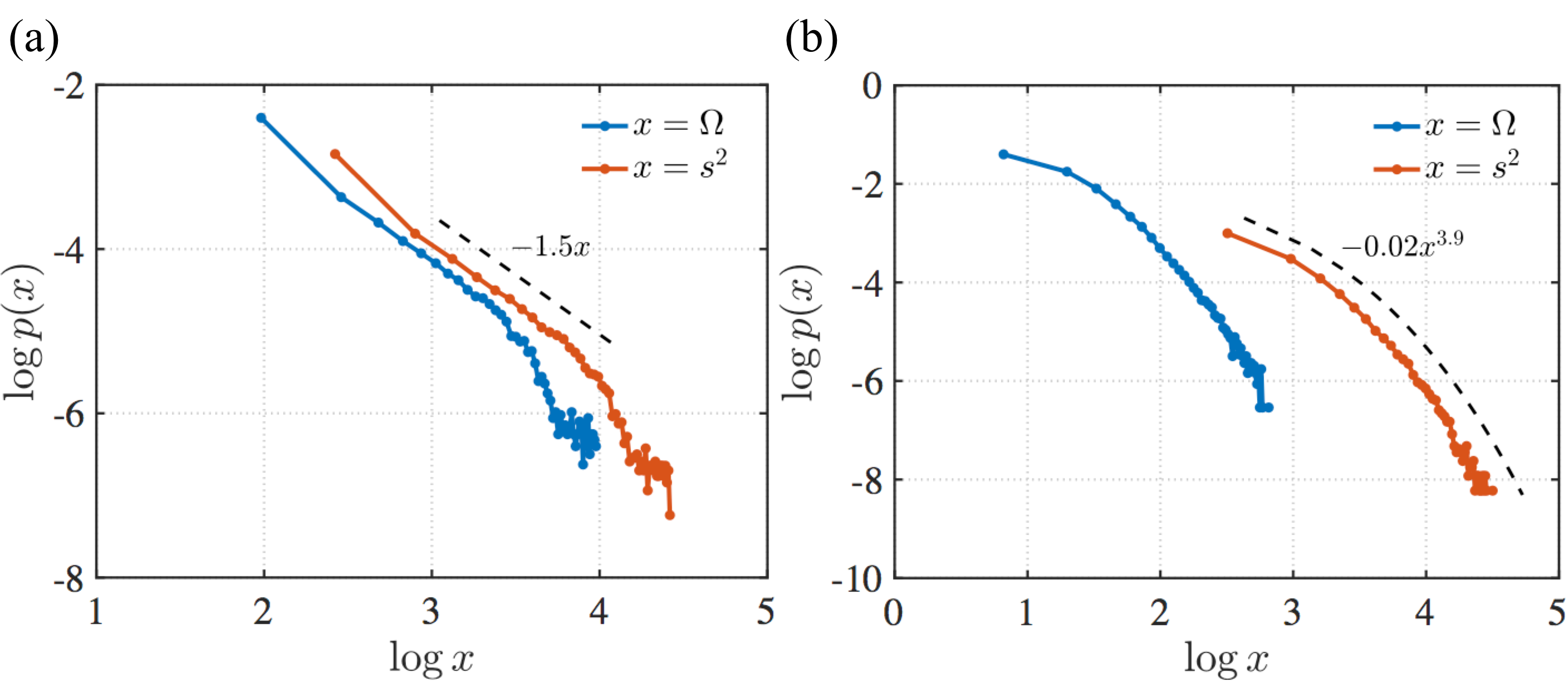}}
  \caption{Enstrophy and node strength (squared) distributions of two- (a) and three-dimensional (b) isotropic turbulence.}
\label{f:str_enst2D3D}
\end{figure}

Let us first characterize the interactions amongst the vortical elements in two- and three-dimensional isotropic turbulence. Following equation~\ref{e:str_enstrophy}, we evaluate the node strength and enstrophy distributions of the flow fields at an instant in time, presented in figure~\ref{f:str_enst2D3D}. The node strength-enstrophy relations from equation~\ref{e:str_enstrophy} is shown here for isotropic turbulence. \cite{benzi1987statistical} found a power-law profile for the enstrophy distribution of two-dimensional decaying isotropic turbulence. The distribution $p(s^2)$ also follows a power-law behavior as observed in figure~\ref{f:str_enst2D3D}~(a). \cite{Taira:JFM16} analyzed the network structure of two-dimensional decaying isotropic turbulence and found the undirected node strength (average of in- and out-strength) distribution $p(s)$ to follow a scale-free behavior if the energy spectrum follows the $k^{-3}$ profile.

For three-dimensional isotropic turbulence, the enstrophy distribution follows a stretched-exponential profile, $p(\Omega) \propto \exp(-a_\Omega \Omega^b)$ \citep{donzis2008dissipation}. We observe a stretched exponential profile for $p(s^2)$ with the same exponent $b$ as that of the enstrophy distribution, as shown in figure~\ref{f:str_enst2D3D}~(b). The difference of $p(s^2)$ for two- and three-dimensional flows can be attributed to the components of vorticity. For three-dimensional turbulence, vorticity is spread over wide scales of structures due to vortex stretching and tilting, which are absent in two-dimensional flows.

\begin{figure}
  \centerline{\includegraphics[width=0.97\textwidth]{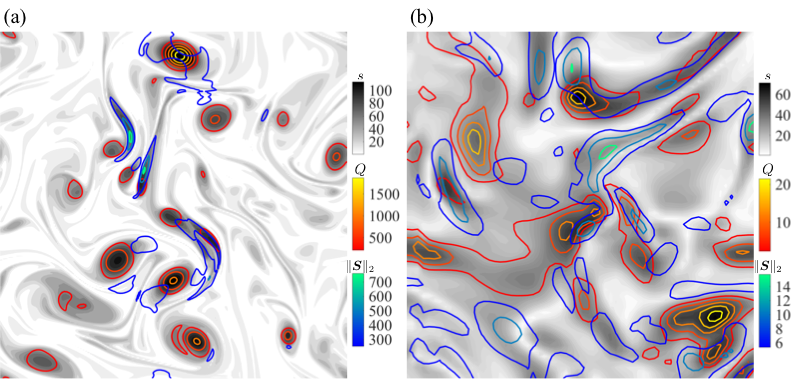}}
  \caption{Comparing vortical structures with high node strength (gray contours), $Q$-criterion (red-yellow contours), and strain (blue-green contours) for (a) two- and (b) three-dimensional isotropic turbulence. Only a slice is shown for (b).}
\label{f:str_QS_contour_2D3D}
\end{figure}

The node strength distributions can be used to highlight vortical elements with high node strength, as shown in figure~\ref{f:str_QS_contour_2D3D}. Isocontours of node strength, positive $Q$-criterion, and magnitude of strain rate tensor $\|\boldsymbol{S}\|_2$, for the two- and three-dimensional flow fields are shown. For both flows, the high node strength regions align with those of high positive $Q$-criterion (vortex core) and $\|\boldsymbol{S}\|_2$ (high shear regions). The node strength-enstrophy relation attributes to the alignment of the node strength with the $Q$ and $\|\boldsymbol{S}\|_2$ measures in physical space. The observations show that the network-based node strength can indeed identify strong vortex cores and shear-layers in turbulent flows. 

\begin{figure}
  \centerline{\includegraphics[width=0.97\textwidth]{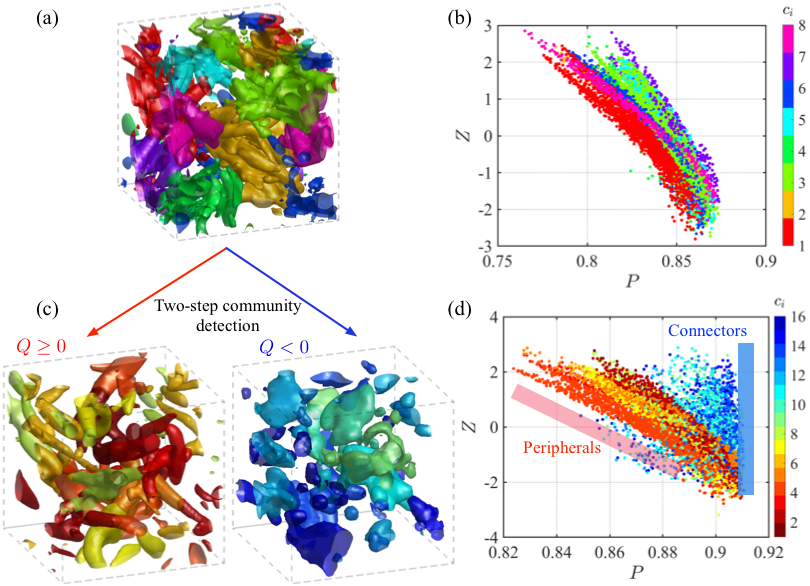}\quad}
  \caption{(a) Community detection in a three-dimensional isotropic turbulence and (b) the corresponding $P-Z$ map. (c) Two-step community detection and (d) the corresponding $P-Z$ map. Connector and peripheral communities have distinct distributions in the $P-Z$ map.}
\label{f:community_split}
\end{figure}

Let us now use the community-based framework to extract vortical communities in isotropic turbulence and identify influential regions to modify the flow using inter- and intra-community strength measures. A demonstration of the community detection algorithm and the corresponding $P-Z$ map applied to a three-dimensional flow field are shown in figure~\ref{f:community_split}~(a) and (b). Here, the initial community detection procedure coarsely identifies regions in the flow field. A clear distinction between connectors and peripherals is not observed using the $P-Z$ map. The continuous nature of the flow field and vortical structures being spatially close to each other can make it challenging for the community detection algorithm to extract distinct vortical structures. Similar observations were made in a previous study for laminar wakes \citep{Meena:PRE18}.

We aid the community detection algorithm by decomposing the flow field into nodes with $Q>0$ and $Q<0$. This is portrayed in figure~\ref{f:community_split}~(c). The communities are identified independently for the two networks. The difference in the number of communities compared to the first step is due to the value of $\gamma_\text{M}$ used, which was found to be similar for both the steps for the given $Re_\lambda$. The new community labels are used to evaluate the $P-Z$ map for the full adjacency matrix, as shown in figure~\ref{f:community_split}~(d). The two-step community detection procedure reveals the communities to broadly follow two correlations in the $P-Z$ map. The communities on the left side of the map possess a negative correlation in the $P-Z$ feature space and the nodes on the left side exhibit no significant correlation. We classify the former as peripheral and later as connector communities. The first group predominantly contains nodes with $Q>0$, the second group with $Q<0$. Thus, most peripheral and connector communities resemble vortex core and shear-layer type structures, respectively. Vortex cores that are spatially isolated in the flow can attribute to their classification as peripherals. Whereas, most shear-layer type structures being located amongst vortical structures make them connectors. We observe similar results for two-dimensional isotropic turbulence and for other cases at various $Re_\lambda$. Given the classification of vortical structures into connector and peripheral communities, we can now identify the important communities and analyze their influence on the flow field.

\subsection{Community-based flow modification}\label{ss:iso_turb_perturbation}

\begin{figure}
  \centerline{\includegraphics[width=0.97\textwidth]{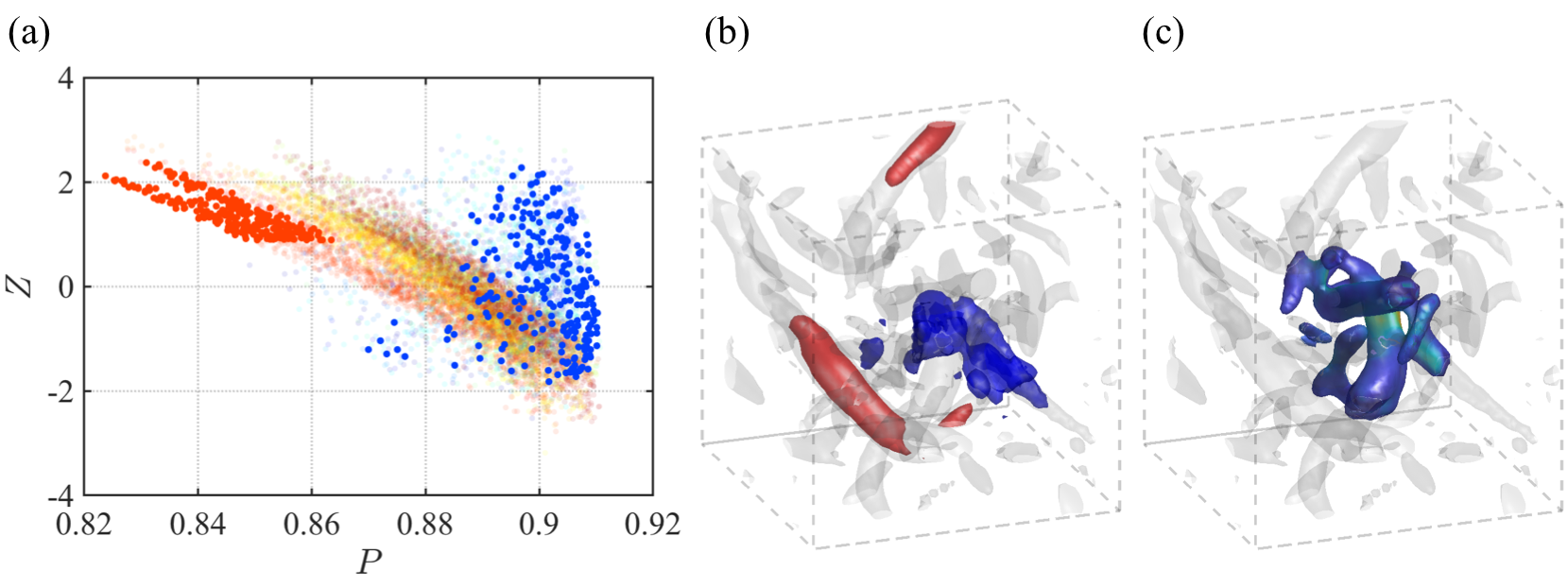}\quad}
  \caption{(a)~Identifying the connector and peripheral in the $P-Z$ map and (b)~the structures in physical space. (c)~Local region around the connector structure.}
\label{f:perturbation_setup}
\end{figure}

We compute the $\overline{P_k}$ of each community $k$ to quantify the strength of influence. The connector and peripheral communities are the ones with $\max_k \overline{P_k}$ and $\min_k \overline{P_k}$, respectively, as highlighted in figure~\ref{f:perturbation_setup}~(a). The dominant nodes of the communities are determined based on $Z$. The peripheral community corresponds to the vortical structure visualized in red and the connector community is comprised of low circulation multi-vortical structures with both strain and rotational regions visualized in blue in figure~\ref{f:perturbation_setup}~(b). 

We compare the influence of the connector structure to modify the flow with the strongest vortex tube and shear-layer based regions, which are known to cause high flow modifications. The vortex tube and shear-layer structures are identified by large amplitudes of $Q>0$ and $Q<0$, respectively, herein denoted as $Q^+$ and $Q^-$. We do not present cases for the peripheral structures as they usually align with $Q^+$.

We add perturbations to the identified influential structures and track the changes to the flow. Impulse perturbations at discrete time $n_\text{t}\Delta t$ are added to the velocity field with time step $\Delta t$ and $n_\text{t} = 0,1,2,\dots$. The perturbed velocity field at time $t$ is given by $\boldsymbol{u}(\boldsymbol{r},t) + \tilde{\boldsymbol{u}}(\boldsymbol{r},\boldsymbol{r}^*,t)$, where $\tilde{\boldsymbol{u}}(\boldsymbol{r},\boldsymbol{r}^*,t) =  \alpha\tilde{\boldsymbol{f}}(\boldsymbol{r},\boldsymbol{r}^*,t)$, $\alpha$ is the amplitude of the perturbation added to the perturbation $\tilde{\boldsymbol{f}}(\boldsymbol{r},\boldsymbol{r}^*,t)$, and $\boldsymbol{r}^*$ is the location of the influential structure. The perturbation is given by
\begin{eqnarray}
        \boldsymbol{f}(\boldsymbol{r},\boldsymbol{r}^*,t) = \frac{\hat{\boldsymbol{e}}_{\boldsymbol{u}(\boldsymbol{r},t)}}{\sqrt{2\pi\Delta V^2}} \sum\limits_{i=1}^{n_\text{p}} \exp\left( \frac{-\| \boldsymbol{r} - \boldsymbol{r}^*_i \|_2^2}{2\Delta V^2} \right) \delta(t - n_\text{t}\Delta t),
\end{eqnarray}
where $\hat{\boldsymbol{e}}_{\boldsymbol{u}(\boldsymbol{r},t)}$ is the unit vector in the direction of $\boldsymbol{u}(\boldsymbol{r},t)$, $\Delta V$ is the volume of the vortical node (grid size), and $\boldsymbol{r}^*_i$ are the locations of the $n_\text{p}$ perturbed nodes. We normalize $\boldsymbol{f}(\boldsymbol{r},\boldsymbol{r}^*,t)$ to give $\tilde{\boldsymbol{f}}(\boldsymbol{r},\boldsymbol{r}^*,t)$ such that $\int_V\|\tilde{\boldsymbol{f}}(\boldsymbol{r},\boldsymbol{r}^*,t)\|_2^2\text{d}V = 1$. A prescribed forcing energy $E$ of
\begin{equation}
    E = \frac{\int\limits_V \| \tilde{\boldsymbol{u}}(\boldsymbol{r},\boldsymbol{r}^*,t) \|_2^2 \text{d}V}{\int\limits_V \|\boldsymbol{u}(\boldsymbol{r},t) \|_2^2 \text{d}V}
\end{equation}
is used to compute the amplitude of perturbation $\alpha$.

We first analyze the influence of the structures with a single perturbation at the initial time. The perturbation amplitude $\alpha$ is chosen such that $E=0.04$, which is a reasonable magnitude for control. Based on the observations, we then employ multiple pulses. The influential structures are tracked in time for given isocontours (or isosurfaces) of $Q$-criterion and are subjected to impulse perturbations with time step $\Delta t/\tau_e(t_0) =1$ and $0.2$ for two- and three-dimensional flows, respectively. We have chosen the time step for these turbulent flows based on the single perturbation analysis to obtain significant flow modification. The difference in time step for two- and three-dimensional isotropic turbulence is to consider the distinct predictability horizon of small-scale motion in the flows \citep{metais1986statistical,lesieur1996new,machiels1997predictability}. While the two-dimensional flow would have predictability horizons spanning over a few eddy turn-over times, the present three-dimensional flow has a shorter time horizon.

We concentrate on the flow evolution around the neighbourhood of the perturbed structures, as shown in figure~\ref{f:perturbation_setup}~(c). The peripheral-based or $Q^+$ perturbations increases the circulation of the vortex core. In contrast, modification of the behavior of multiple vortical structures by the connector-based and $Q^-$ perturbations enhances local mixing in the flow, as we discuss below. Based on the observations from the present study (for which the influential structures are identified only at initial time), we have also performed preliminary analyses with the structure identification procedure repeated as the flow evolves. We observe similar results showing perturbation of the connector structures significantly enhancing turbulent mixing compared to perturbing $Q^+$ and $Q^-$ structures. In the present study, we concentrate on the structures identified at initial time to characterize the influence of the network-based structures on turbulence.
Mixing enhancement in turbulent flows plays a key role in various engineering applications. Mixing enhancement by stirring have been attributed to generation of shear dominated filaments in the flow field \citep{spencer1951mixing,aref1984stirring,ottino1990mixing,eggl2018gradient}. Here, we analyze the effect of the perturbations to enhance local mixing in isotropic turbulence based on the present network-based framework. We use fluid particle tracking to quantify local mixing in the flow field. To quantify mixing, we consider the use of two-species fluid tracking \citep{coppola2001nonlinear}. The first species is initialized one integral length scale around the centroid of the perturbed structure, the region previously shown in figure~\ref{f:perturbation_setup}~(c), and the second over the rest of the flow field. Given the velocity field $\boldsymbol{u}(\boldsymbol{r},t)$, the time evolution of the fluid particle at $\boldsymbol{r}_p$ is given by
\begin{equation}
    \frac{\text{d}\boldsymbol{r}_p}{\text{d}t} = \boldsymbol{u}(\boldsymbol{r}_p,t).
\end{equation}
A second-order accurate Runge--Kutta scheme is implemented for time integration \citep{yu2012studying}.

Local mixing is quantified by measuring the information entropy using the two species of particles in the domain \citep{kang2004colored,cookson2019efficiently}. The flow field is discretized into cells and the entropy of the two species of particles is evaluated for each cell. The total information entropy of the whole domain at an instant in time is given by
\begin{equation}
    S = - \sum_{i=1}^{n_\text{c}}\left [  w_i \sum_{k=1}^{2} \left( n_{i,k} \log{n_{i,k}} \right) \right ],
\end{equation}
where $n_\text{c}$ is the number of cells with which the full domain is discretized, $n_{i,k}$ is the number of particles of $k$th species in $i$th cell, and $w_i$ is the weight factor. If the $i$th cell contains no particles or only particles of a single species, $w_i=0$, else $w_i=1$. Moreover, the relative entropy measure $\kappa$ \citep{kang2004colored} is given by
\begin{equation}
	\kappa(t) = \frac{S(t) - S(t_0)}{S_{\text{max}} - S(t_0)},
\end{equation}
where $S(t_0)$ is the entropy for initial particle distribution and $S_{\text{max}}$ is the maximum possible entropy increase over the domain, which is achieved when each cell contains equal number of particles of each species. For each perturbation setup, we normalize $\kappa$ by the baseline $\kappa_{\text{base}}$ as $\tilde{\kappa}(t) = \left[ \kappa(t) - \kappa_{\text{base}}(t) \right] / \| \kappa_{\text{base}}(t) \|_{\infty}$, thus measuring the mixing enhancement compared to the baseline flow. The baseline flow field is initialized with particles in the same pattern as each perturbed simulation to evaluate the respective $\kappa_{\text{base}}$.

\begin{figure}
  \centerline{\includegraphics[width=0.97\textwidth]{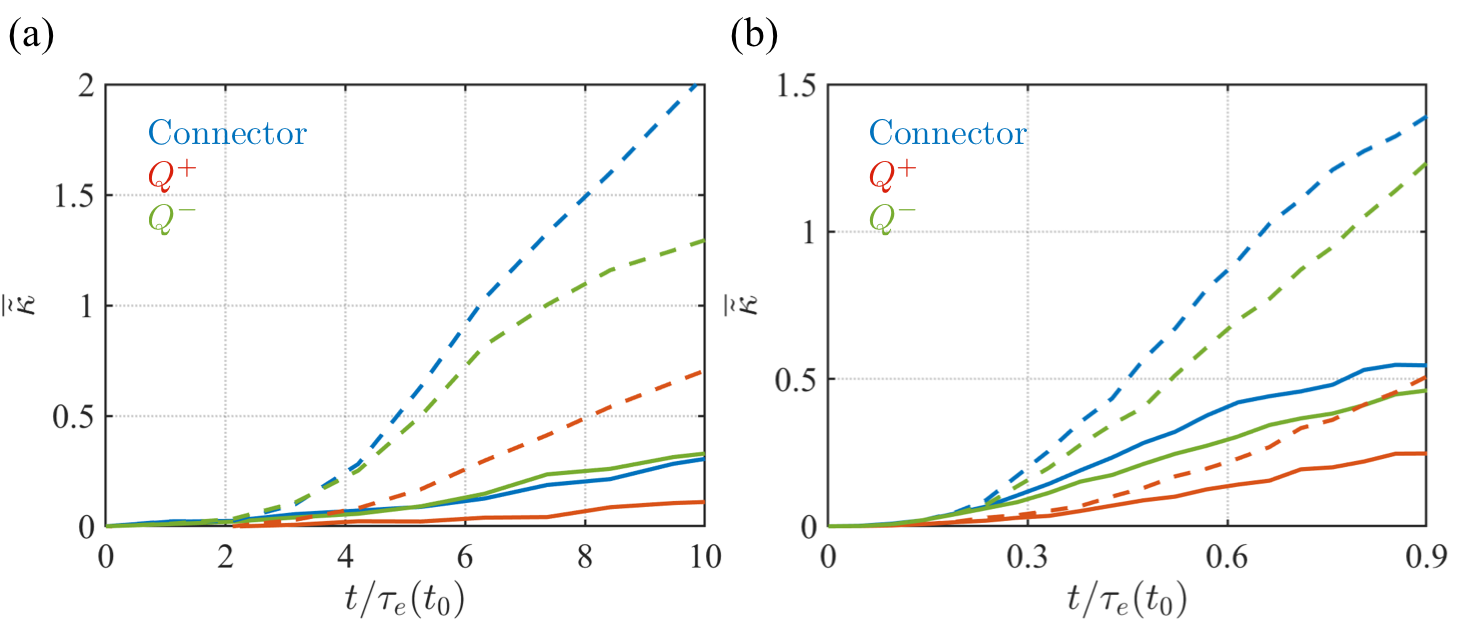}}
  \caption{Ensemble average of normalized relative particle entropy in time for (a) two- and (b) three-dimensional isotropic turbulent flows subjected to connector-based, $Q^+$, and $Q^-$ perturbations. Lines --- denote results for single perturbation and $--$ for multiple perturbation. Ensemble is computed using 10 cases.}
\label{f:mixing_enh_single_sus_2D3D}
\end{figure}

Ensemble averages of $\tilde{\kappa}$ for connector, $Q^+$, and $Q^-$-based perturbations performed on a number of two- and three-dimensional isotropic turbulent flows are shown in figure~\ref{f:mixing_enh_single_sus_2D3D}. For the two-dimensional isotropic turbulence with a single perturbation at initial time, the connector-based perturbation achieves mixing enhancement similar to that of $Q^-$. Both connector-based and $Q^-$ perturbations outperforms $Q^+$ for local mixing enhancement. With multiple pulses, connector-based perturbation generates entropy two times compared to the baseline flow, which is $54\%$ more than the $Q^-$ perturbations. The perturbation using $Q^+$ just leads to strengthening of vortex cores without significant spreading of particles, which will be visualized shortly.

Using the results of two-dimensional flows as a guideline, we perform the analysis on three-dimensional flows, as shown in figure~\ref{f:mixing_enh_single_sus_2D3D}~(b). A single pulse add to the connector at initial time generates $\tilde{\kappa}$ values $0.55$ times that of the baseline flow, which is $20\%$ more than $Q^-$ perturbation. With multiple pulses, connector-based perturbations increase the entropy by $1.4$ times compared to the baseline flow, which is $17\%$ more than the $Q^-$ perturbations. Multiple pulses of $Q^+$ perturbation under-performs compared to even the single connector-based perturbation. According to the studies on mixing enhancement using stirrers \citep{aref1984stirring,eggl2018gradient}, we expect the $Q^-$ perturbation to achieve better mixing. The present results are in agreement with these studies. 

\begin{figure}
  \centerline{\includegraphics[width=0.8\textwidth]{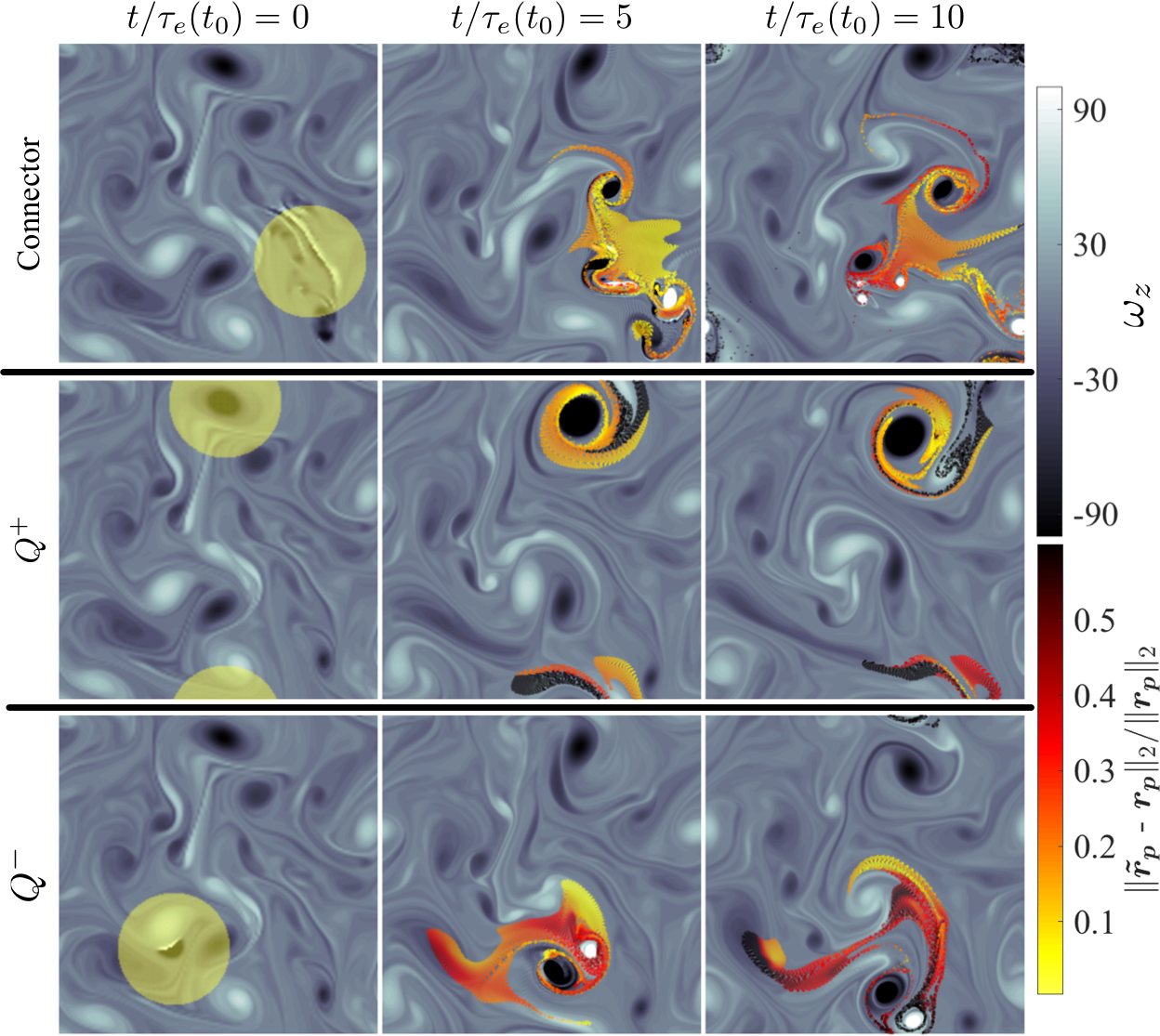}}
  \caption{Time evolution of a two-dimensional isotropic turbulent flow subjected to multiple perturbations. Fluid tracers initialized around the perturbation are coloured by the change in trajectory compared to the corresponding baseline trajectory.}
\label{f:mixing_flow_2D_sus}
\end{figure}

With the above observations quantifying the effectiveness of connectors to modify turbulence and enhance local turbulence mixing, let us analyze the flow fields. We consider the time evolution of a two-dimensional isotropic decaying turbulent flow subject to multiple pulses of connector-based, $Q^+$, and $Q^-$ perturbations, as shown in figure~\ref{f:mixing_flow_2D_sus}. The fluid particle species initialized one integral length around the perturbation is shown. The yellow-red colour gradient represents the amount of change in trajectories compared to the corresponding baseline trajectory. A uniform colour distribution indicates effective mixing. The connector-based perturbation achieves the narrowest range in colour distribution at the final instant. Whilst $Q^+$ and $Q^-$ perturbations achieve higher magnitudes of change in trajectories for certain particles, highlighted in black, some particles are spread out least, as highlighted in bright yellow.

Multiple pulses on the $Q^+$ structure strengthens the vortex core, making a large distinct vortex. The tracers rotate around the vortex core due to the vorticity. Only the tracers at the boundary of the vortex are spread out significantly. The $Q^-$ perturbation, comprising of the strained region between the vortex-dipoles, forms a jet-like flow. With multiple perturbations, the jet-like flow spreads the tracers more effectively compared to that achieved in the baseline and $Q^+$ perturbations. These observations are in agreement with recent findings that vortex-dipole like structures in two-dimensional isotropic turbulence promote effective flow modification \citep{jimenez2020monte}. The connector structures are comprised of a long shear-layer type vortical structure and near-by vortices of low vorticity. Multiple perturbations of the shear-layer type structure lead to the formation of small scale vortices, which enhances flow mixing compared to the baseline as well as $Q^+$ and $Q^-$ perturbations.

\begin{figure}
  \centerline{\includegraphics[width=0.8\textwidth]{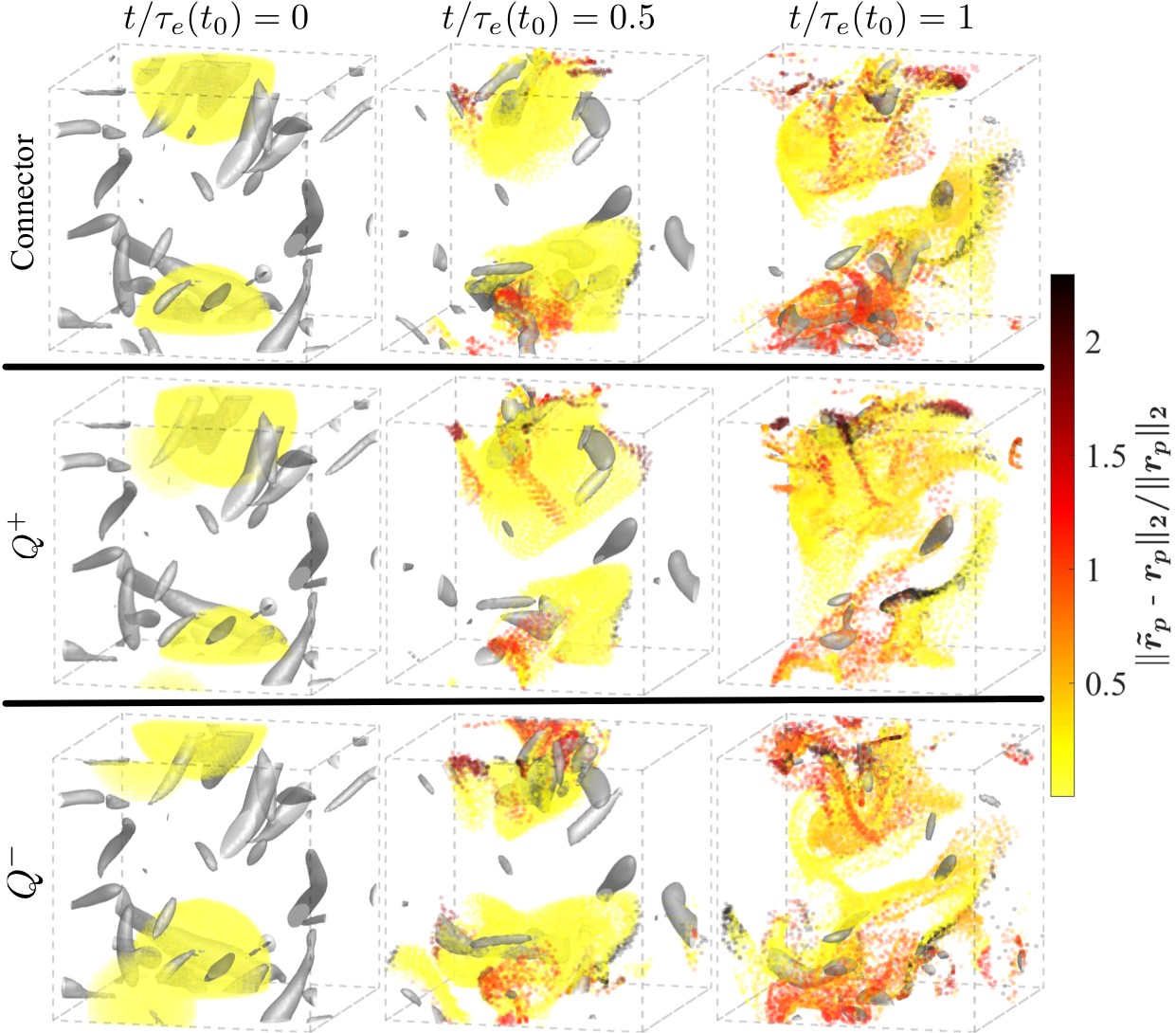}}
  \caption{Time evolution of a three-dimensional isotropic turbulent flow subjected to multiple perturbations. Vortical structures are depicted by isosurface of $Q$-criterion. Fluid tracers initialized around the perturbation are coloured by the change in trajectory compared to the corresponding baseline trajectory.}
\label{f:mixing_flow_3D_sus}
\end{figure}

The evolution of particle tracers in a three-dimensional turbulent flow subjected to multiple perturbations on connector, $Q^+$, and $Q^-$  structures are presented in figure~\ref{f:mixing_flow_3D_sus}. Here, the $Q^+$, $Q^-$, and connector structures are spatially located close to each other at the initial time. The flow field at the final instant corresponding to connector-based perturbation has the narrowest range in colour distribution of the tracers, depicting highest mixing enhancement. The broadest range in colour distribution corresponds to $Q^+$ perturbation, which denotes the least mixing enhancement achieved. Multiple pulses of connector-based perturbations lead to the generation of small scale structures. On the other hand, the vortical structures decay in time with $Q^+$ and $Q^-$ perturbations.

\begin{figure}
  \centerline{\includegraphics[width=0.7\textwidth]{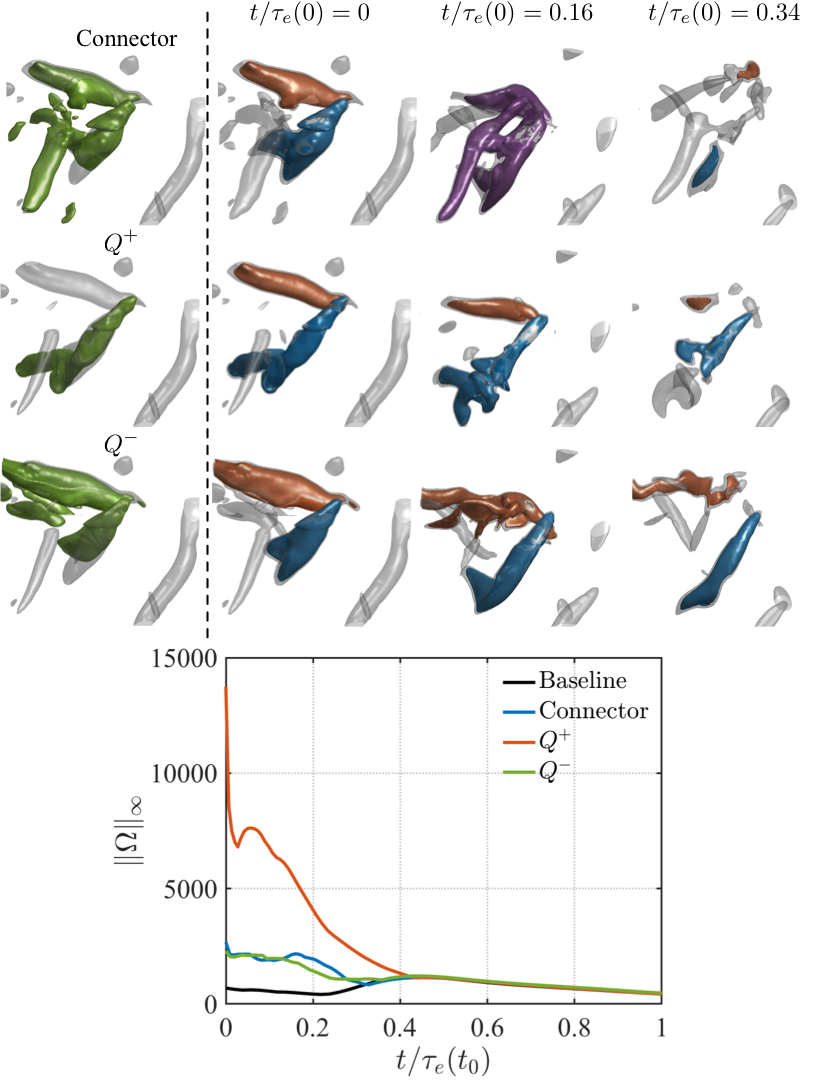}}
  \caption{Modification of two vortex tubes in a three-dimensional isotropic turbulent flow using $Q^+$, $Q^-$, and connector-based perturbations. The connector effectively modifies both the vortex cores as shown using the isosurface of $Q$-criterion and local growth of maximum enstrophy $\|\Omega\|_\infty$ between time $t/\tau_e(0) = 0.12$ and $0.2$.}
\label{f:struct_modi_3D_run1}
\end{figure}

Next, let us examine the local flow region around the perturbation to analyze the effect of the network-based connector to modify its neighbouring vortical structures. We analyze the time evolution of the local flow region around the connector, $Q^+$, and $Q^-$ structures, as presented in figure~\ref{f:struct_modi_3D_run1}. The vortical structures are tracked using the isosurface of $Q$-criterion, with a constant value of $Q$ in time. A single pulse is added to the connector, $Q^+$, and $Q^-$ structures at the initial time. The perturbed structures at the initial time are shown on the left column, visualized by the green isosurfaces. The neighbouring vortex tubes under consideration in each case (row) are depicted by the blue and red isosurface. The growth of maximum enstrophy, $\|\Omega\|_\infty$, is also evaluated to observe modification of the vortex cores \citep{foias1989gevrey,ayala2017extreme}. We note that at time $t/\tau_e(0) = 0.34$, the connector structure leads to modification of both the neigboring vortex tubes and formation of smaller scale structures. These small-scale structures induce enhanced local spreading of the fluid particles, as observed in figure~\ref{f:mixing_flow_3D_sus}.

Perturbation of the $Q^+$ structure (shown in blue) leads to the increase in circulation of the vortex core and eventual break-up of the blue isosurface at time $t/\tau_e(0) = 0.16$. The modification of the $Q^+$ structure is also observed by the peak of $\|\Omega\|_\infty$ at $t/\tau_e(0) = 0.05$. The curved geometry of the $Q^+$ vortex core at the initial time may instigate instabilities when perturbation is added, resulting in the significant modification of the vortex core. Nonetheless, the $Q^+$ perturbation has negligible influence on the neighbouring vortex highlighted in red. Note that these observations are similar to the characteristics of a network-based peripheral structure. The $Q^-$ structure is the strongest shear-layer structure, which is located between the two neighbouring vortex tubes. Thus, the $Q^-$ perturbation results in the modification of both the blue and red isosurfaces. Both the vortex cores decay in time with no peaks appearing for $\|\Omega\|_\infty$ over time, suggesting lower modification of the vortex cores achieved by the $Q^-$ structure compared to $Q^+$ perturbation.

The connector structure is comprised of the shear-layer region between the two neighbouring vortex tubes and an adjacent vortical structure with low circulation. This adjacent vortical structure merely decays within a short time in the $Q^+$ and $Q^-$ simulations. With connector-based perturbation, the adjacent vortical structure connects with the two neighbouring vortex tubes through the shear-layer, forming one large vortical structure highlighted in purple at time $t/\tau_e(0) = 0.16$. Eventually, the connection results in the significant break-up of both the blue and red isosurfaces, leading to the formation of smaller scales of vortical structures. The observations show the advantage of connectors to modify multiple vortical structures compared to $Q^+$ perturbations. The local growth of $\|\Omega\|_\infty$ between $t/\tau_e(0) = 0.13$ and $0.2$  denotes effective modification of the vortex cores compared to that achieved by $Q^-$ perturbation. The connection amongst the vortical structures, in the form of vorticity, is not aligned with the direction of rotation of the two neighbouring vortex tubes, which may result in instabilities, leading to significant break-up of the vortex cores. The above illustrative examples demonstrate the ability of network-based methodologies to extract vortical structures of low circulation that can effectively influence neighbouring vortical structures in turbulent flows.

\section{Concluding remarks}\label{s:conclusion}

A network community-based formulation was introduced to extract flow modifying vortical structures in two- and three-dimensional isotropic turbulence. The network framework considered the vortical elements in a flow as nodes of a graph and captured the web of interactions amongst them, quantified by the induced velocity, as the weighted edges. The interaction-based framework was used to identify groups of closely connected vortical nodes, called communities. The interactions amongst the communities were used to identify the most influential communities which can modify the system dynamics significantly. Taking advantage of the inter-community interactions, we showed that local turbulent mixing can be enhanced using vortical structures attributed with low circulation. The goal of this network-based framework was not to alter the global turbulent flow, but influence certain key vortical structures under the settings of isotropic turbulence.

We decomposed the governing equation for a networked system using the intra- and inter-community interactions. The strengths of these interactions were used to identify the connector and peripheral nodes, which have the highest and least influence on other communities, respectively. The node with the maximum total interaction strength was identified as the hub. The ability of these influential nodes to modify the networked dynamics was demonstrated on a model fluid flow of a collection of discrete point vortices. Impulse velocity perturbations were added to the connector, peripheral, and hub nodes identified at the initial time. The connector community effectively modified trajectories of the other communities compared to the hub and peripheral-based perturbations. We then applied the community-based formulations to identify influential structures in two- and three-dimensional isotropic turbulence. The connector and peripheral structures were found to resemble shear-layer and vortex core type structures, respectively. Adding perturbations to the connector structures, which have low vorticity, led to enhanced local flow mixing compared to the effect of perturbing the strongest shear-layer and vortex tube.

Within the current work, we have not addressed the necessary computational effort for performing the network-based analysis.  However, there are emerging network-based characterization \citep{bai2019randomized} and modeling techniques \citep{Nair:JFM15} for fluid flows that can be leveraged to perform the present analysis with reduced computational resource.  With these developments, the present characterization and open-loop control offers a pathway for future closed-loop flow control efforts to modify the dynamics of vortical structures residing in a complex turbulent flow field.

\section*{Acknowledgements}

The authors thank the US Army Research Office (Grants: W911NF-17-1-0118 and W911NF-19-1-0032, Program Manager: Dr.~Matthew J.~Munson) for supporting this work. Some of the simulations were performed using the computational resource made available by the US Department of Defense High Performance Computing Modernization Program. We acknowledge the thought provoking discussions with Professors Steven L. Brunton, James C.~McWilliams, Drs.~Aditya G.~Nair, and Chi-An Yeh.

\appendix

\section{Reynolds number and dimension reduction independence of network measures}\label{app:str_grid_Re}

\begin{figure}
  \centerline{\includegraphics[width=0.97\textwidth]{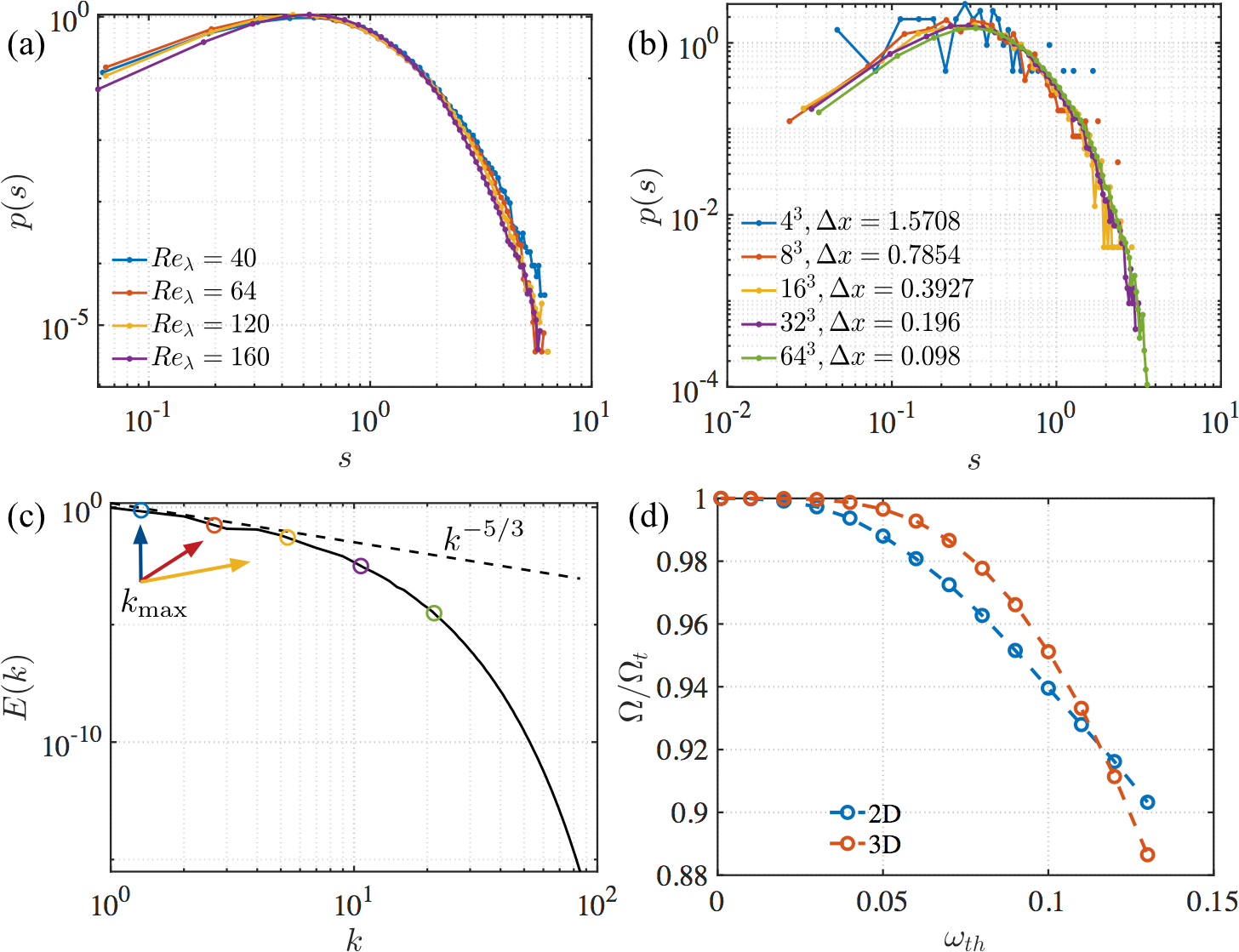}\quad}
  \caption{(a) Collapse of node strength distribution with increase in Reynolds number. (b) Convergence of node strength distribution over sampling rate in three-dimensional turbulence of $Re_\lambda = 40$. (c) Maximum resolvable wave number for various grid resolution or sampling rate. (d)Variation of enstrophy with vorticity threshold in two- and three-dimensional turbulence.}
\label{f:appendix_DimRe}
\end{figure}

We assess the effect of Reynolds number on the network representation of vortical interactions in isotropic turbulence by evaluating the node strength distribution for three-dimensional flow fields with various $Re_\lambda$, as shown in figure \ref{f:appendix_DimRe}~(a). The strength distributions of flows with different Reynolds numbers collapse to a single curve with the use of appropriate non-dimensionalization for the edge weights. The results suggest that the network formulation is able to capture the key characteristics of the universal scaling of isotropic turbulence. The fine resolution of spatial discretization required to simulate turbulent flows also pose a challenge to analyze the interactions amongst the vortical elements. We subsample the flow field to reduce the dimension of the flow field being analyzed.

The effect of sub-sampling the turbulent flow field on the network measures is evaluated by comparing the node strength distribution of various low resolution flow fields. A three-dimensional flow field of $Re_\lambda = 40$ with a grid resolution of $256\times256\times256$ is used. We sub-sample the data to various smaller grid resolutions and compute the node strengths, as shown in figure \ref{f:appendix_DimRe}~(b). The node strength computed based on the non-dimensionalized edge weight allows scaling of the distributions. The node strength distributions collapse into a single distribution and converge with higher grid resolution. The $k_{\text{max}}$ for each low-resolution grid used in the node strength distribution computation is depicted in the energy spectra in figure \ref{f:appendix_DimRe}~(c). The grid resolution using a $64\times64\times64$ grid satisfies the $k_{\text{max}} \eta \ge 1$ condition. The results also demonstrate the application of network-based formulation from a discrete to continuous description of the flow field.

We only analyze vortical elements in the flow field with high vorticity, extracted through vorticity threshold. The choice of the vorticity threshold of $\omega_{th}=0.05||\boldsymbol{\omega}(\mathbf{r})||_\infty$ is given by analyzing the effect of thresholding on the total enstrophy of the flow field, as shown in figure \ref{f:appendix_DimRe}~(d). For both two- and three-dimensional turbulence, $\omega_{th}=0.05||\boldsymbol{\omega}(\mathbf{r})||_\infty$ results in retaining more than $98\%$ of the enstrophy of the original flow field. This is similar to the values used to analyze structures of high vorticity \citep{jimenez1993structure,moisy2004geometry}, capturing the influential regions of the flow.\\

\bibliography{refs}
\bibliographystyle{jfm}

\end{document}